\documentclass[preprint,nofootinbib,12pt]{revtex4}

\usepackage{amsmath}
\usepackage{amsthm}
\usepackage{amsfonts}
\usepackage{graphicx}
\usepackage{psfrag}
\usepackage{verbatim}

\newcommand{\GeV}{{\rm GeV}}

\newcommand{\beq}{\begin{equation}}
\newcommand{\eeq}{\end{equation}}
\newcommand{\beqa}{\begin{eqnarray}}
\newcommand{\eeqa}{\end{eqnarray}}

\newcommand{\lsim}{\mathrel{\rlap{\lower4pt\hbox{\hskip1pt$\sim$}}
    \raise1pt\hbox{$<$}}}         
\newcommand{\gsim}{\mathrel{\rlap{\lower4pt\hbox{\hskip1pt$\sim$}}
    \raise1pt\hbox{$>$}}}         
\newcommand{\eq}[1]{(\ref{#1})}

\theoremstyle{definition} \newtheorem{definition}{Definition}
\theoremstyle{plain} \newtheorem{theorem}{Theorem}

\begin{document}


\vspace*{.0cm}

\title{Parameter counting in models with global symmetries}

\author{Joshua Berger}\email{jb454@cornell.edu}
\author{Yuval Grossman}\email{yuvalg@lepp.cornell.edu}

\affiliation{\vspace*{4mm}Institute for High Energy Phenomenology\\
Newman Laboratory of Elementary Particle Physics\\
Cornell University, Ithaca, NY 14853, USA\vspace*{6mm}}


\vspace{1cm}
\begin{abstract}
We present rules for determining the number of physical parameters in
models with exact flavor symmetries. In such models the total number
of parameters (physical and unphysical) needed to described a matrix
is less than in a model without the symmetries. Several toy examples
are studied in order to demonstrate the rules. The use of global
symmetries in studying the minimally supersymmetric standard model
(MSSM) is examined.
\end{abstract}

\maketitle

\section{Introduction}
When modeling a physical system, it is important to understand the
relationship between the symmetries in the model and the number of
physical parameters involved.  Consider for example a hydrogen atom in
a uniform magnetic field.  Before turning on the magnetic field, the
hydrogen atom is invariant under spatial rotations, which are
described by the $SO(3)$ group.  Furthermore, there is an energy
eigenvalue degeneracy of the Hamiltonian: states with different
angular momenta have the same energy.  This degeneracy is a
consequence of the symmetry of the system.

When magnetic field is added to the system, it is conventional to pick
a direction for the magnetic field without a loss of
generality. Usually, we define the positive $z$ direction to be the
direction of the magnetic field. Consider this choice more carefully.
A generic uniform magnetic field would be described by three real
numbers: the three components of the magnetic field.  However, the
magnetic field breaks the $SO(3)$ symmetry of the hydrogen atom system
down to an $SO(2)$ symmetry of rotations in the plane perpendicular to
the magnetic field.  The one generator of the $SO(2)$ symmetry is the
only valid symmetry generator now; the remaining two $SO(3)$
generators in the orthogonal planes are broken.  These broken symmetry
generators allow us to rotate the system such that the magnetic field
points in the $z$ direction:
\beq\label{h-magnetic}
  O_{xz} O_{yz} \begin{pmatrix} B_x \\ B_y
  \\ B_z \end{pmatrix} = \begin{pmatrix} 0 \\ 0
  \\ B_z^\prime \end{pmatrix}, 
\eeq 
where $O_{xz}$ and $O_{yz}$ are rotations in the $xz$ and $yz$ planes
respectively. The two broken generators were used to rotate away two
unphysical parameters, leaving us with one physical parameter, the
value of the magnetic field. That is, all measurable quantities in the
system depend only on one new parameter, rather than the na\"ive
three.  In addition, the broken symmetry lifts the degeneracy of the
energy eigenvalues.

The results described above are more generally applicable.
Particularly, they are useful in studying the flavor physics of
quantum field theories.  Consider a gauge theory with matter content.
This theory always has kinetic and gauge terms, which have a certain
global symmetry $G_f$ on their own.  However, in adding a potential,
which consists of a linear combination of all renormalizable operators
that respect the imposed symmetries, the global symmetry may be broken
down to a smaller symmetry group.  In breaking the symmetry, there is
an added freedom to rotate away unphysical parameters, as when a
magnetic field is added to the hydrogen atom system.  In order to
analyze this process, we define a few quantities.  The added potential
has coefficients that can be described by $N_\mathrm{general}$
parameters in a general basis.  The global symmetry $H_f$ of the
entire model has fewer generators than $G_f$ and we call the
difference in the number of generators $N_\mathrm{broken}$.  Finally,
the quantity that we would ultimately like to determine is the number
of parameters affecting physical measurements, $N_\mathrm{phys}$.
These numbers are related by the well-known
rule~\cite{Santamaria:1993ah} (for a review see, for example,
Ref.~\cite{Nir:2007})
\beq \label{countrule}
N_\mathrm{phys} = N_\mathrm{general} - N_\mathrm{broken}.  
\eeq 
Furthermore, this rule applies separately for both real parameters
(masses and mixing angles) and phases.  A general, $n \times n$
complex matrix can be parametrized by $n^2$ real parameters and $n^2$
phases.  Imposing restrictions like Hermiticity or unitarity reduces
the number of parameters required to describe the matrix.  A Hermitian
matrix can be described by $n(n+1)/2$ real parameters and $n(n-1)/2$
phases, while a unitary matrix can be described by $n(n-1)/2$ real
parameters and $n(n+1)/2$ phases.

The rule given by \eq{countrule} can be applied to the standard model.  We
consider only terms involving fermions, stating results for the Higgs
field when they are relevant.  The Yukawa potential for the interactions in
terms of the quark $SU(2)_L$ doublet, $Q_L$, the lepton $SU(2)$
doublet, $L_L$, the $SU(2)_L$ singlet fields, $U_R$, $D_R$, $E_R$, and
the Higgs doublet, $H$, is
\beq \label{smpotential}
V = Y^U_{ij} (\overline{Q_L})_i (U_R)_j H + Y^D_{ij} (\overline{Q_L})_i (D_R)_j
\tilde{H} + Y^E_{ij} (\overline{L_L})_i (E_R)_j \tilde{H} + \mathrm{h.c.},
\eeq
where $Y^F$ are $3 \times 3$ complex matrices in a general basis.  We
use $\tilde{H} = \epsilon H^*$, where $\epsilon$ is the anti-symmetric
matrix in $SU(2)_L$ space.

The interactions in this sector are parametrized by three complex $3
\times 3$ matrices, which contain a total of 54 parameters (27 real
parameters and 27 phases) in a general basis.
These parameters also break a large global symmetry of the kinetic and
gauge terms in the model down to the familiar baryon number and lepton family
number symmetries of the full standard model, 
\beq
U(3)_Q \times U(3)_U \times U(3)_D \times U(3)_L \times U(3)_E 
\to U(1)_B \times U(1)_e \times U(1)_\mu \times U(1)_\tau. 
\eeq
While $U(3)^5$ has $45$ generators, the remaining symmetry
group has only $4$ and thus $N_\mathrm{broken}=41$. This broken symmetry
allows us to rotate away a large number of the parameters by moving to
a more convenient basis.  Using \eq{countrule}, the number of physical
parameters should be given by
\beq 
N_\mathrm{phys} = 54 - 41 = 13.
\eeq
In addition, there are the three gauge couplings, the two Higgs
parameters and the strong CP phase for a total of 19 parameters in the
standard model. These parameters can be split into real parameters
and phases.  The five unitary matrices generating the symmetry of the
kinetic and gauge terms have a total of 15 real parameters and 30
phases and the symmetry is broken down to a symmetry with only four
phase generators.  Thus,
\beq
N^{(r)}_\mathrm{phys} = 27 - 15 = 12,\qquad N^{(i)}_\mathrm{phys} = 27 - 26
= 1.
\eeq
We interpret this result by saying that of the 12 real parameters, 9
are the fermion masses and three are the CKM matrix mixing angles.
The one phase is the CP-violating phase of the CKM mixing matrix.

In studying new models, it is particularly important to properly count
the number of parameters. The number of physical parameters is, in
principle, the number of measurements required in order to fully
determine a model.  Once these measurements are made, it should be
possible to test the model with all further measurements.  The
standard model is so successful because all the parameters have been
measured to some extent and further measurements have verified
significant predictions of the model to high precision.  The current
parametrization appears to be sufficient to describe the quark sector
at scales below $100~\GeV$~\cite{Amsler:2008}. The failure of the SM
parametrization in the lepton sector have been used as indicators of
new lepton flavor physics~\cite{Amsler:2008}.

In this paper, we extend the rule for parameter counting to theories
where global symmetries are imposed on the potential terms.  In
particular, we consider cases where part of the flavor symmetry
present in the kinetic and gauge terms is restored.  In section 2, a
rule for analyzing these cases is presented.  Simple toy examples are
discussed to highlight the use of the rule.  In section 3, the rule is
applied to studying global symmetry constraints in the MSSM.  The
results of imposing symmetries are compared to the constrained MSSM (cMSSM).

\section{Rules for Parameter Counting}\label{rules}

In general, we distinguish between two ways in which one could impose
a global symmetry. The symmetry can be imposed on the whole model, or
only on a specific sector.  Clearly, a symmetry of a specific sector
is broken by higher order terms. Yet, in terms of parameter counting we
care about the tree level parameters. For example, in the SM the
custodial symmetry is respected only by the Higgs sector and it is
broken at one loop.

In the following, we study both cases and show that the general result
is the same: the total number of parameters, $N_{\rm total}$, needed
to describe a model in a general basis is reduced compared to a model
without such symmetries. The specific number of parameters needed in
each case is different.

The most general type of terms on which we consider imposing a global
symmetry has the form 
\beq \label{interaction-form} 
Y_{ij} \phi^{(1)}_i \phi^{(2)}_j \dots, 
\eeq 
where $\phi^{(1)}$ and $\phi^{(2)}$ have $n$ generations each, $Y$ is
an $n \times n$ mixing matrix and $\dots$ represents other
(flavor-singlet) factors that ensure that the term is a gauge group
singlet.  Multiple terms of the form \eq{interaction-form} may be
present.  It is therefore possible that some of the symmetries imposed
could hold for some terms, but be broken explicitly by others.
Furthermore, if one or more of the gauge symmetries of the model is
broken, then it is possible to allow the imposed symmetries to be
broken by the gauge sector.  

We start by looking at a simple toy model. Consider the leptonic
sector of the standard model, but with an imposed $SU(2)$ symmetry
such that two of the lepton masses are the same. Since the leptonic
Yukawa matrix can be diagonalized without breaking any gauge
symmetry, if the symmetry is imposed on the Yukawa sector, it will
hold for the entire model.  Thus, the cases of imposing the symmetry
on the model and on the Yukawa sector only are the same for this
choice of matter content.  The only interaction term that it is
necessary to consider for now is the third term in
\eq{smpotential}, $Y^E_{ij} (\overline{L_L})_i (E_R)_j \tilde{H}$.
As we show below, the result is that the total number of parameters
required to describe this term in an arbitrary basis, $N_{\rm total}$,
is reduced from 18 to 15.

In an arbitrary basis, we begin to decompose the matrix $Y^E$, first
performing a polar decomposition~\cite{Knapp:1996}:
\beq \label{polardecomp}
Y^E = R \Phi,
\eeq
where $R$ is Hermitian with positive eigenvalues and $\Phi$ is
unitary.  The next step is to perform a spectral decomposition on
$R$:
\beq \label{spectraldecomp}
Y^E = U^\dagger D U \Phi,
\eeq
where $U$ is unitary and $D = \mathrm{diag}(m_e, m_e, m_\tau)$ (recall
that we choose $m_e=m_\mu$). Clearly, $U$ can be taken to have unit
determinant in general. The final step is to apply a Cartan
decomposition~\cite{Knapp:1996} on $U$.  The involution of choice here
will allow us to break $U$ into the product of a matrix in $U(2)
\times U(1)$ and a matrix generated by the remaining generators of
$SU(3)$.  At this point, to illustrate the general procedure, we
explicitly perform steps outlined in the Appendix.  The Cartan
decomposition theorem (see the Appendix for a statement of the theorem
and more details) then allows us to write
\beq \label{cartandecomp}
U = k \exp (\mathbf{p}),
\eeq
where $k \in U(2) \times U(1)$ and $\mathbf{p} = \sum_{j = 4}^7 i a_j
\lambda_j/2$, $a_j$ are real numbers and $\lambda_i$ are the Gell-Mann
matrices.  Note that $\mathbf{p}$ is described by 4 parameters, the
$a_j$. The final form of the matrix $R$ is then
\beqa \label{finalform}
R &=& \exp (-\mathbf{p}) \begin{pmatrix}U_{2\times2}^\dagger & 0\\ 0 &
e^{-i\alpha}\end{pmatrix} \begin{pmatrix}m_e & 0 & 0\\ 0 & m_e & 0\\ 0
& 0 & m_\tau\end{pmatrix} \begin{pmatrix}U_{2\times2} & 0\\0 &
e^{i\alpha}\end{pmatrix} \exp (\mathbf{p})\nonumber \\ &=& \exp (
-\mathbf{p})\begin{pmatrix}m_e &0 & 0\\0 & m_e& 0\\ 0 & 0 &
m_\tau\end{pmatrix} \exp(\mathbf{p}).
\eeqa
The main result following from \eq{finalform} is that it only 6
parameters are required to describe the matrix $R$ in this way. They
are the two eigenvalues and the 4 $a_j$. This is in contrast to the
usual 9 for a general $3\times 3$ Hermitian matrix.

The decomposition of $R$ given by \eq{finalform} demonstrates the fact
that the value of $N_\mathrm{general}$ is reduced when symmetries
are imposed.  In this case, the usual 18 is decreased to 15, of which
7 are real parameters and 8 are phases.  As a check, the symmetry
breaking pattern is
\beq \label{symmetrybreaking}
U(3)_L \times U(3)_E \to U(2)_e \times U(1)_\tau.
\eeq
Thus, there are 13 broken symmetry generators, $N_\mathrm{broken}=13$.
Using the fact that $N_\mathrm{general}=15$ and using
Eq.~\eq{countrule} we get
\beq
N_\mathrm{phys} = N_\mathrm{general} - N_\mathrm{broken}=15-13=2.
\eeq
Indeed there are two flavor parameters in this model, $m_e$ and $m_\tau$.

Now consider a more general model with one term of the form in
\eq{interaction-form}, $Y_{ij} \phi^{(1)}_i \phi^{(2)}_j$.  Without
any restrictions, $2 n^2$ parameters would be required to describe $Y$
in a general basis.  Whenever symmetries are imposed on such terms,
this number is reduced.  The degeneracies of the matrix eigenvalues
ensure that one can always parametrize the matrix with fewer
parameters than one would na\"ively expect.  As a first step in
proving the general formula, consider imposing an $n_1$-fold
eigenvalue degeneracy on $Y$, with $1 < n_1 \leq n$.  Since $Y$ can be
diagonalized, this is equivalent to imposing an $SU(n_1)$ symmetry.
Using results obtained in the Appendix, the required number of
parameters is reduced by $n_1^2 - 1$ and thus $N_\mathrm{general} =
2n^2 - n_1^2 + 1$ out of which, $n^2 + 1 - n_1(n_1+1)/2$ are real and
$n^2 - n_1(n_1-1)/2$ are phases. 

With this result for an imposed $SU(n_1)$ symmetry, it is possible to
iteratively extend the symmetry group to $SU(n_1) \times \dots \times
SU(n_k)$.  For each imposed $SU(n_j)$, $n_j^2 -1$ parameters can be
removed.  Thus, the most general result for an $n$ general model with
two $n$ fields transforming in the (anti-)fundamental of the imposed
symmetry group is
\beq
N_\mathrm{phys} = 2n^2 - \sum_{j = 1}^k (n_j^2 - 1).
\eeq
In terms of real parameters, $N^{(r)}_{\rm phys}$, and phases,
$N^{(i)}_{\rm phys}$, the result is
\beq \label{full-degen-rules}
N_\mathrm{phys}^{(r)} = n^2 - \sum_{j = 1}^k \left(\frac{n_j(n_j +
  1)}{2} - 1\right),\qquad N_\mathrm{phys}^{(i)} = n^2 - \sum_{j = 1}^k \frac{n_j(n_j-1)}{2}.
\eeq

Some complications arise when more terms are added to the potential,
particularly when one field appears in multiple potential terms.  The
cases of a full model symmetry and a sector symmetry cease to be the
same as the interaction matrices cannot always be diagonalized
concurrently with the gauge interactions.  The case of a sector
symmetry is trivial to extend.  In this case, the symmetry must hold
if the interaction matrices in the sector were diagonalized.  In this
diagonal basis, a certain number of eigenvalues need to be degenerate
in order for the symmetry to be manifest.  The case of an interaction
matrix with degenerate eigenvalues was discussed above and applies
also to this case. In particular, the rule \eq{full-degen-rules} apply
to each individual interaction matrix.  For model-wide symmetries,
there are correlations between the change of basis matrices allowed in
different terms.  We demonstrate a general procedure for determining
the correlations below.

Consider a model with three fields $\phi^{(k)}$ that have $n$
generations each.  Suppose further that the non-gauge interaction
terms have the form
\beq
L = Y^{(2)}_{ij} \phi^{(1)}_i \phi^{(2)}_j \cdots + Y^{(3)}_{ij}\phi^{(1)}_i
\phi^{(3)}_j \cdots.
\eeq
A typical example of a part of a model with interactions of this form
is the quark-sector Yukawa interactions in the standard model.  An
$SU(n_1) \times \dots \times SU(n_k)$ symmetry is imposed with all
fields having their first $n_1 + \dots + n_k$ components transform in
the fundamental.  Na\"ively, one might expect the number of parameters
to be simply twice that of the one-interaction-term model with the
same symmetry.  However, there is a reduction in the number of
parameters due to the fact that the change of basis matrix $U$ in
\eq{spectraldecomp} must be the same for both Yukawa matrices in order
for the symmetry to hold in some basis.  Of the physical parameters
subtracted off in \eq{full-degen-rules}, $\sum_j (n_j - 1)$ were real
eigenvalues that are now degenerate and $\sum_j n_j(n_j-1)/2$ real
parameters and phases were parameters in $U$.  Thus, since $n$ of the
phases of $U$ always multiply out independent of the symmetry, the $U$
matrix has the same number of real parameters and phases
\beq
{n(n-1) \over 2} - \sum_{j=1}^k {n_j(n_j-1) \over 2}.
\eeq
Thus, we count twice the number of parameters as in the one term case,
then subtract off the number of parameters in each repeated $U$
matrix.  Using this counting, we find that the total number of
parameters required is
\beq \label{twotermcount}
N_\mathrm{general}^{(r)} = \frac{n(3n+1)}{2} - \sum_{j = 1}^k
\frac{(n_j + 4)(n_j - 1)}{2},\qquad
N_\mathrm{general}^{(i)} = \frac{n(3n+1)}{2} - \sum_{j = 1}^k
\frac{n_j(n_j-1)}{2}.
\eeq
If the symmetry is only required to hold in the Yukawa sector, but may
be broken by the weak interactions, then there really are twice as
many parameters in this case as in the case with one
interaction term.  That is
\beq\label{twotermweak}
N_\mathrm{general}^{(r)} = 2n^2 - 2\sum_{j = 1}^k \left(\frac{n_j(n_j +
  1)}{2} - 1\right),\qquad N_\mathrm{general}^{(i)} = 2n^2 - 2\sum_{j = 1}^k \frac{n_j(n_j-1)}{2}.
\eeq
Finally, if we demand only that the first term has such a symmetry,
but allow the symmetry to be broken by the other term, then only
$Y^{(2)}$ is restricted.  In a general basis, we subtract off the
parameters of $U$ that are unnecessary for that matrix
\beq\label{twotermdown}
N_\mathrm{general}^{(r)} = 2n^2 - \sum_{j = 1}^k \left(\frac{n_j(n_j +
  1)}{2} - 1\right),\qquad N_\mathrm{general}^{(i)} = 2n^2 - \sum_{j = 1}^k \frac{n_j(n_j-1)}{2}.
\eeq
Any other model can be handled by accounting for the appropriate
relation among the $U$ matrices, described in one of the
cases \eq{twotermcount}, \eq{twotermweak} or \eq{twotermdown}.

\section{Parameter Counting in the MSSM}

Even with imposed R-parity, the MSSM has 124 parameters, which is much
more than the 19 of the standard
model~\cite{Dimopoulos:1995,Haber:2001}.  In order to make any
specific, quantitative predictions using the model, it is necessary to
make some assumptions about the flavor structure of the model.  One of
the most popular models that does so is the constrained MSSM (cMSSM),
which has only 4 new parameters and one undetermined sign.  The cMSSM
involves a number of arbitrary assumptions about the parameters that
appear in the low-energy Lagrangian.  A different approach is to start
imposing symmetries on the interactions at some UV scale, which we can
then run down to the scales being studied.  In order to see how this
approach works and how the rules derived in section \ref{rules} help
us in studying the MSSM, we consider a toy version of the MSSM.

The toy model has only two generations of quarks, no leptons and exact
R-parity.  The superpotential for quark multiplets is
\beq
W  = Y^U_{ij} Q_i U_j H_u + Y^D_{ij} Q_i D_j H_d,
\eeq
where $Y^Q_{ij}$ are $2\times 2$ complex matrices.  See for
example~\cite{Haber:2001} for the choice of conventions for
representations under the MSSM gauge group.  The SUSY-breaking
potential for the squarks is given by
\beq\label{toy-susy-breaking}
V_\mathrm{soft} = (A^U_{ij} \tilde{Q}_i \tilde{U}_j H_u + A^D_{ij}
\tilde{Q}_i \tilde{D}_j H_d + \mathrm{h.c.}) + (M^2)^Q_{ij}
\tilde{Q}_i^\dagger \tilde{Q}_j + (M^2)^U_{ij} \tilde{U}_i^\dagger
\tilde{U}_j + (M^2)^D_{ij} \tilde{D}_i^\dagger\tilde{D}_j,
\eeq
where $A^Q_{ij}$ are complex $2\times 2$ matrices, and $(M^2)^F_{ij}$
are Hermitian $2\times 2$ matrices.  

Before restricting the model, we compute the number of flavor
parameters in this toy MSSM.  There are four $2\times 2$ complex
matrices and three $2\times 2$ Hermitian matrices, which in the
absence of symmetries gives the counting
\beq\label{mssm-no-sym-gen}
N_{\rm general}^{(r)} = 25, \qquad N_{\rm general}^{(i)} = 19.
\eeq
The full $U(2)^3$ flavor symmetry is broken by the interaction terms
\beq\label{mssm-no-sym-break}
U(2)^3 \to U(1)_B.
\eeq
Using \eq{countrule}, we then find that
\beq\label{mssm-no-sym-phys}
N_{\rm phys}^{(r)} = 22, \qquad N_{\rm phys}^{(i)} = 11.
\eeq
The non-supersymmetric model with the same gauge and matter content
has only 5 real parameters in the quark sector.

As in the non-supersymmetric case, there are a number of ways to
impose a symmetry.  Obviously, we could require the symmetry to hold
through all sectors of the model.  However, the symmetry could also be
imposed on the SUSY-breaking sector and broken by the SUSY sectors.
It could be imposed on the two potentials, but broken by weak
interactions.  Finally, it could be imposed on the up quarks only, but
broken by the down quarks or vice versa.

Consider the various ways of imposing a $U(1)$ symmetry on the lighter
generation of quarks.  This symmetry will automatically guarantee a
second $U(1)$ for the heavy quarks.  The least restrictive ways to
impose the symmetry are to demand either that it hold only for the up
quarks or only in the soft SUSY-breaking potential.  It turns out that
both scenarios have the same number of parameters.  In the case where
symmetry is imposed only on the up quark matrices, the only
restriction is that all the up quark interaction matrices be
simultaneously diagonalizable.  If the matrices are written in the
form \eq{spectraldecomp}, then all their $U$ and $\Phi$ matrices must
be the same up to an overall diagonal phase matrix.  The down
interaction matrices are not affected by this restriction.  If the
symmetry is imposed for both types of quarks, but only in $V_{\rm
  soft}$, then all the $U$ matrices must be the same within the
SUSY-breaking sector.  Both cases lead to the counting:
\beq\label{mssm-u1soft-gen}
N_{\rm general}^{(r)} = 21, \qquad N_{\rm general}^{(i)} = 15.
\eeq
Since the imposed symmetry is broken by other sectors, the symmetry
breaking is
\beq\label{mssm-u1soft-break}
U(2)^3 \to U(1)_B.
\eeq
With \eq{countrule}, it is then easy to see that the number of
physical parameters is given by
\beq\label{mssm-u1soft-phys}
N_{\rm phys}^{(r)} = 18, \qquad N_{\rm phys}^{(i)} = 7.
\eeq

The number of parameters is further reduced if we demand that the
symmetry hold for both potentials and for both types of quarks.  Not
only are the $U$ matrices now correlated, but so are the $\Phi$
matrices.  The number of parameters in a general basis is
\beq\label{mssm-u1pot-gen}
N_{\rm general}^{(r)} = 18, \qquad N_{\rm general}^{(i)} = 12.
\eeq
There is no additional symmetry for the full model, and we count that
\beq\label{mssm-u1pot-phys}
N_{\rm phys}^{(r)} = 15, \qquad N_{\rm phys}^{(i)} = 4.
\eeq

The next more restrictive case is imposing the $U(1)$ throughout the
model.  Progressing to this case is as simple as extending the
correlations from the previous cases to the entire model, so that
\beq\label{mssm-u1-gen}
N_{\rm general}^{(r)} = 17, \qquad N_{\rm general}^{(i)} = 11.
\eeq
The extra $U(1)$ symmetry now holds on the model so part of the flavor
symmetry is restored
\beq\label{mssm-u1-break}
U(2)^3 \to U(1)_u \times U(1)_c.
\eeq
Thus, the number of physical parameters is given by
\beq\label{mssm-u1-phys}
N_{\rm phys}^{(r)} = 14, \qquad N_{\rm phys}^{(i)} = 4.
\eeq

Next, we study models where we impose minimal flavor violation
(MFV) on the Yukawas and their supersymmetry-breaking extensions.  MFV
is defined in the spurion formalism by saying that the only
flavor-violating spurions are the standard model Yukawa matrices.  To
leading order, this forces $A^F = a^F Y^F$, where $a^F$ is a complex
number, and $(M^2)^F = (m^2)^F 1$, where $(m^2)^F$ is a real number.
The parameter counting in the SUSY-breaking sector is as
follows. There are two additional parameters for each three-scalar
coupling and one extra for each mass.  In the end, we find that in a
general basis
\beq\label{mssm-mfv-gen}
N_{\rm general}^{(r)} = 13, \qquad N_{\rm general}^{(i)} = 10.
\eeq
Only baryon number is left after breaking the symmetry, so
that
\beq\label{mssm-mfv-phys}
N_{\rm phys}^{(r)} = 10, \qquad N_{\rm phys}^{(i)} = 2.
\eeq

The cMSSM is a restriction of the MFV case.  It is assumed that at
some high scale all the scalar masses are equal, all the three-scalar
couplings $a^F$ are equal and all the gaugino masses are equal.  In
the full model with leptons, these restrictions hold between baryons
and leptons as well.  Furthermore, the new interactions are assumed to
be CP-conserving so that there are no new CP violating physical
phases.  With these conditions, the counting in a general basis is
\beq\label{cmssm-gen}
N_{\rm general}^{(r)} = 10, \qquad N_{\rm general}^{(i)} = 8,
\eeq
so that in the physical basis
\beq\label{cmssm-phys}
N_{\rm phys}^{(r)} = 7, \qquad N_{\rm phys}^{(i)} = 0.
\eeq
The non-supersymmetric analogue of this model had only 5 flavor
parameters: four quark masses and a mixing angle.  Thus, there are two
new flavor parameters here which we can take to be the SUSY-breaking
squark mass $m^2_0$ and the triscalar coupling $a_0$.  These two
additional parameters in the quark sector, together with the
SUSY-breaking Higgs parameter $b$ and the gaugino mass $m_{1/2}$, are
the only new parameters.  The superpotential mass parameter $\mu$
can be related to the Higgs VEV and is not counted as new.  An extra
undetermined sign comes from moving to a more convenient
parametrization where the Higgs parameters $\mu$ and $b$ are traded
for $m_Z$ and $\tan \beta$.  The two sets of parameters contain the
same information up to the sign of $\mu$ which is not fixed by
fixing $m_Z$ and $\tan \beta$.  This ambiguity arises from the
fact that the scalar Higgs potential of the MSSM depends only on
$|\mu|^2$ and not on $\mu$. 

Most of the counting outlined in \eq{mssm-no-sym-gen}-\eq{cmssm-phys}
above extends trivially to constraining the full MSSM.  The main
complication is the additional generation in the fermion sectors.  The
additional generation allows an $SU(2)$ flavor symmetry to be imposed.
This type of symmetry can then be handled using the rules derived in
section \ref{rules}.  Maintaining exact R-parity, the superpotential
for the fermion multiplets is given by
\beq \label{superpotential}
W = Y^U_{ij} Q_i U_j H_u + Y^D_{ij} Q_i D_j H_d + Y^L_{ij} L_i E_j H_d,
\eeq
where $Y^F_{ij}$ are $3\times 3$ complex matrices and $\mu$ is a
complex number.  The SUSY-breaking potential for the fields in these
multiplets is given by:
\begin{multline} \label{susy-breaking}
V_\mathrm{soft} = (A^U_{ij} \tilde{Q}_i \tilde{U}_j H_u + A^D_{ij}
\tilde{Q}_i \tilde{D}_j H_d + A^L_{ij} \tilde{L}_i \tilde{E}_j H_d +
\mathrm{c.c.}) + \\ 
(M^2)^Q_{ij} \tilde{Q}_i^\dagger \tilde{Q}_j +
(M^2)^U_{ij} \tilde{U}_i^\dagger\tilde{U}_j + (M^2)^D_{ij}
\tilde{D}_i^\dagger\tilde{D}_j + (M^2)^L_{ij}
\tilde{L}_i^\dagger\tilde{L}_j + (M^2)^E_{ij}
\tilde{E}_i^\dagger\tilde{E}_j,
\end{multline}
where $A^F_{ij}$ are complex $3\times 3$ matrices and $(M^2)^F_{ij}$ are
Hermitian $3\times 3$ matrices.  

The results of the parameter counting for various imposed symmetries
are described in table \ref{symmetry-models}.  The four columns show
the number of real and imaginary parameters in a general basis and in
the physical basis for the potential of the chiral flavored fields.
The first row gives the counting for the case when no symmetry is
imposed.  This is the MSSM-124 model.  The second through fourth lines
describe the case where only $U(1)$ family symmetry is imposed.  On
the second line, the symmetry is broken by the superpotential.  On the
third line, it is broken by weak gauge interactions.  On the fourth
line, it holds through all renormalizable terms in the model.  The
fifth through seventh lines describe the case where $SU(2)$ is imposed
with the first two generations transforming as a doublet and the third
as a singlet.  The same three symmetry-breaking possibilities are
presented.  On the eighth line, we present the case of MFV where only the
leading term in powers of the Yukawa  matrices is kept.  Finally, the
case with maximal $SU(3)$ flavor symmetry is presented.

\begin{table}[!t]
\begin{center}
\begin{tabular}{c c c c c c}
\hline
Imposed Symmetry & Broken By &$N_{\rm general}^{(r)}$ & $N_{\rm general}^{(i)}$ & $N_{\rm
  phys}^{(r)}$ & $N_{\rm phys}^{(i)}$\\
\hline
None & ~ & 84 & 69 & 69 & 41\\
Fermion Family & SUSY Interaction & 66 & 51 & 51 & 23\\
~ & Weak Interactions & 51 & 36 & 36 & 10\\
~ & All & 48 & 33 & 33 & 9\\
$SU(2)$ Flavor & SUSY Interaction & 56 & 49 & 41 & 21\\
~ & Weak Interactions & 37 & 30 & 23 & 6\\
~ & All & 35 & 28 & 22 & 6\\
Leading MFV & ~ & 35 & 30 & 20 & 4\\
$SU(3)$ Flavor & ~ & 20 & 21 & 11 & 3\\
\hline
\end{tabular}
\caption{Parameter counting in the chiral multiplet potentials of the
  MSSM with various imposed symmetries in the potentials only and in
  the entire Lagrangian for the model.  The large $SU(N)$ symmetries
  are necessarily broken, possibly
  spontaneously~\cite{Barbieri:1996}.}\label{symmetry-models}
\end{center}
\end{table}

\section{Conclusions}

It is clear that the Standard Model is a low energy description of a
more fundumental theory.  The introduction of new states and
symmetries into the Lagrangian adds many new interaction matrices.
The hierarchy of the SM Yukawa matrices as well as the new physics
flavor puzzle~\cite{Nir:2007} motivate the idea that new flavor
symmetries or approximate symmetries could exist in more fundamental
interactions.  If such symmetries exist, then parameter counting
may be non-trivial.  The number of parameters required in a general
basis is less than if the symmetries were not imposed.  We derived
rules for accounting for this reduction in the number of parameters.
We demonstrated the analysis for a series of toy models, leading up to
counting the number of parameters in the MSSM with various imposed
flavor symmetries.  The results obtained for the MSSM are summarized
in Table \ref{symmetry-models}.  However, the methods used above are
general and can be used to study other potential UV completions of the
Standard Model.

\section*{Acknowledgments} 
We thank Yossi Nir and Jo\~{a}o Silva for helpful discussions. This
research is supported by the NSF grant PHY-0355005.

\appendix
\section{Cartan Decomposition of a Unitary Matrix}

The Cartan decomposition theorem is a theorem about semisimple Lie
groups that gives a decomposition for elements of the group.  In all
the cases that we consider, the matrix we would like to decompose is
an element of the semisimple Lie group $U(n)$.  It is trivial to
factor out the overall phase of such a matrix, and thus we
consider below the decomposition of a matrix $U \in SU(n)$ for
simplicity.

The mathematical definitions and theorems can be found, for example,
in~\cite{Knapp:1996}. The specific decomposition process is inspired
by the work of~\cite{Divakaran:1980}.  In order to understand the idea
of a Cartan decomposition, we need to make some definitions.
\begin{definition}
Let $\mathfrak{g}$ be a semisimple Lie algebra.  An automorphism
$\theta$ of $\mathfrak{g}$ with square equal to the identity is called
an involution.  An involution is a Cartan involution if the symmetric
bilinear form
\begin{equation}
B_\theta(X, Y) = - B(X, \theta Y)
\end{equation}
is positive definite, where $B$ is the Killing form of $\mathfrak{g}$.
\end{definition}
The second definition is slightly technical, but for practical
purposes, the involutions we use satisfy this condition.  For more
details, please see~\cite{Knapp:1996}.  Since $\theta^2 = 1$, $\theta$
has eigenvalues $\pm 1$ on $\mathfrak{g}$.  Thus, we can decompose
$\mathfrak{g} = \mathfrak{l} \oplus \mathfrak{p}$, where
$\mathfrak{l}$ and $\mathfrak{p}$ are the eigenspaces of $\theta$
corresponding to eigenvalues $+ 1$ and $- 1$ respectively.  This is
the Cartan decomposition on a Lie algebra level.  It is trivial to see
by applying the involution that $[\mathfrak{l}, \mathfrak{l}] =
\mathfrak{l}$; that is, the commutator of any two Lie algebra elements
with eigenvalue 1 under $\theta$ has eigenvalue 1.  This result means
that the eigenspace $\mathfrak{l}$ is actually a Lie subalgebra.
Extending this to the Lie group level is non-trivial and the theorem
is due to Cartan.
\begin{theorem}
Let $G$ be a semisimple Lie group with Lie algebra $\mathfrak{g}$.
Let $\theta$ be a Cartan involution on $\mathfrak{g}$.  Let
$\mathfrak{g} = \mathfrak{l} \oplus \mathfrak{p}$ be the eigenspace
decomposition for $\theta$.  Finally, let $K$ be the subgroup of $G$
with Lie algebra $\mathfrak{l}$.  Then
\begin{enumerate}
\item
there exists a Lie group automorphism $\Theta$ of $G$ with
differential $\theta$ and with $\Theta^2 = 1$,
\item
the subgroup of $G$ that is invariant under $\Theta$ is $K$,
\item
the mapping $K \times \mathfrak{p} \to G$ given by $(k, \mathbf{p}) \mapsto
k\exp (i\mathbf{p})$ is a diffeomorphism.
\end{enumerate}
\end{theorem}
The first consequence can be interpreted as saying that for group
elements infinitesimally different from the identity, the relation
$\Theta(g) = 1 + i \epsilon \theta(\mathbf{g}) + O(\epsilon^2)$ holds.
For any group element, this can be extended to $\Theta(g) =
\exp(i\theta(g))$.  The third consequence is the main result that we
need in order to perform the decomposition.  Effectively, it allows us
to factor an element $g \in G$ into a product of an element $k \in K$
and another element of $SU(n)$ given by $\exp(i\mathbf{p})$ for
$\mathbf{p} \in \mathfrak{p}$ by using the fact that the map defined
in condition 3 is a diffeomorphism and that $\mathfrak{g} =
\mathfrak{k} \oplus \mathfrak{p}$.

Now, consider the group $G = SU(n)$.  Suppose we want to factor an
element $g \in G$ into a product of an element $k$ which is block
diagonal with the first $n_1 \times n_1$ block an element of $SU(n_1)$
for $n_1 < n$ and another element $p \in SU(n)$ whose generators are
all different from those of $k$.  Along similar lines to
\cite{Divakaran:1980}, we choose an involution
\beq \label{involution} 
\theta(\mathbf{g}) = \begin{pmatrix} -1_{n_1} & 0 \\
  0 & 1_{n-n_1}\end{pmatrix} \mathbf{g} 
\begin{pmatrix} -1_{n_1} & 0 \\
  0 & 1_{n-n_1}\end{pmatrix}.
\eeq
This involution is in fact a Cartan involution.  Furthermore, its
eigenspace with eigenvalue $+1$ is all special unitary matrices that
are block diagonal with blocks of size $n_1\times n_1$ and $(n -
n_1)\times(n - n_1)$.  This subalgebra is generated by matrices whose
upper-left block are generators of $SU(n_1)$ and remaining entries are
zero, whose lower-right block are generators of $SU(n - n_1)$ and
remaining entries zero, or which are diagonal phase matrices with
determinant 1.  The orthogonal eigenspace is generated by the
generators whose entries are all off the diagonal block.  By the
Cartan decomposition theorem, we can then write any $SU(n)$ matrix in
the form
\beq
U = \begin{pmatrix} U_{n_1} & 0\\
0 & U_{n - n_1}\end{pmatrix} \exp(i \mathbf{p}),
\eeq
where $U_k$ is a matrix in $U(n_k)$ with $\det U_{n_1} \det U_{n-n_1}
= 1$ and where $\mathbf{p}$ is in the Lie algebra of $SU(n)$ and is
generated by matrices whose diagonal blocks are zero.

Note that this process can be iterated: we can then decompose $U_{n -
 n_1}$ in a similar way.  Ultimately, the matrix $U$ can be written
in the form
\beq\label{full-cartan}
U = \begin{pmatrix} U_{n_1} & ~ & ~\\
~ & U_{n_2} & ~ & ~\\
~ & ~ & \ddots & ~\\
~ & ~ & ~ & U_{n_{k+1}}\end{pmatrix} \exp(i \mathbf{p}),
\eeq
where $n_1 + n_2 + \dots + n_{k+1} = n$, $\det(U(n_1)U(n_2) \dots
U(n_k)) = 1$, and $\mathbf{p}$ is a linear combination of generators
whose entries are all off the diagonal block.  The condition on the
determinants can be removed by allowing $U \in U(n)$ rather than
$SU(n)$.  

The essential result for this work is that a Hermitian matrix with
degenerate eigenvalues can be written in terms of fewer parameters
than if no degeneracy were present.  Let $R$ be a Hermitian matrix
that has $k$ degenerate eigenvalues, with the first one, $r_{n_1}$,
being $n_1$-fold degenerate, the second one, $r_{n_2}$, being
$n_2$-fold degenerate, and so on.  By spectral decomposition, the
matrix can be written as
\beq
R = U^\dagger D U
\eeq
where $D$ is diagonal.  The matrices $U$ are unitary since $R$ is
Hermitian.  Now, decompose $U$ using \eq{full-cartan}.  The
decomposition yields
\beq \label{degen-reduction}
R = \exp(-i \mathbf{p}) U^\dagger D U
\exp(i \mathbf{p}) 
= \exp(-i \mathbf{p}) D \exp(i \mathbf{p}),
\eeq
such that 
\beq
D=
\begin{pmatrix} r_{n_1} 1_{n_1} & ~ & ~ & ~\\
~ & \ddots & ~ & ~\\
~ & ~ &r_{n_k} 1_{n_k} & ~\\
~ & ~ &~ & D_{n_{k+1}}\end{pmatrix}, \qquad
U=\begin{pmatrix} U_{n_1} & ~ & ~ & ~\\
~ & \ddots & ~ & ~\\
~ & ~ &U_{n_k} & ~\\
~ & ~ &~ & P_{n_{k+1}} \end{pmatrix},
\eeq
where $D_{n_{k+1}}$ is diagonal matrix and $P_{n_{k+1}}$ is a diagonal
matrix of (different) phases.  In order to count the number of
parameters necessary to describe this matrix, we can count the number
of parameters in $U$ before performing the reduction of
\eq{degen-reduction} and subtract off the number of parameters removed
by decomposing.  $U$ is an $n\times n$ unitary matrix, which na\"ively
has $n(n-1)/2$ real parameters and $n(n+1)/2$ phases.  By
decomposition, we removed $\sum_j n_j(n_j-1)/2$ real parameters and
$\sum_j n_j(n_j+1)/2 + n - \sum_j n_j$ phases.  In the counting of
the phases, the first sum comes from adding up the parameters in the
unitary matrices $U_{n_j}$ and the second two terms come from adding
up the phases in $P_{n_{k+1}}$.  Furthermore, the number of real
parameters in $D$ is $n - \sum_j (n_j -1)$.  Thus, the number of real
parameters in $R$ is
\beq
N_R^{(r)} = \frac{n(n+1)}{2} - \sum_j \left(\frac{n_j(n_j+1)}{2} - 1\right)
\eeq
and the number of phases in $R$ is
\beq
N_R^{(i)} = \frac{n(n-1)}{2} - \sum_j \left(\frac{n_j(n_j-1)}{2}\right).
\eeq


\end{document}